\documentclass[final,5p,times,twocolumn,numbers]{elsarticle}
\usepackage{amssymb}
\usepackage{lipsum}

\usepackage{amsmath}
\usepackage{xcolor}
\usepackage{hyperref}
\usepackage{ulem}
\usepackage{natbib}
\usepackage{lineno}

\journal{Physics Letters B}

\begin{document}

\begin{frontmatter}

\title{Investigation of the Spectator Effect on Light Nuclei Production in Nucleus-Nucleus Collisions at High Baryon Density Region}

\author[ccnu]{Hongcan Li}

\author[ccnu]{Li'Ang Zhang\corref{cor1}}
\ead{liangzhang@mails.ccnu.edu.cn}

\author[ccnu]{Junyi Han}

\author[ccnu]{Yaping Wang\corref{cor1}}
\ead{wangyaping@ccnu.edu.cn}

\author[ucas]{Junlin Wu}

\author[ucas]{Guannan Xie}

\author[ucas,imp]{Gao-Chan Yong}

\cortext[cor1]{Corresponding author}

\affiliation[ccnu]{organization={Key Laboratory of Quark and Lepton Physics (MOE) and Institute of Particle Physics, Central China Normal University},
city={Wuhan},
postcode={430079}, 
country={China}}
\affiliation[ucas]{organization={School of Nuclear Science and Technology, University of Chinese Academy of Sciences},
city={Beijing},
postcode={101408}, 
country={China}}
\affiliation[imp]{organization={Institute of Modern Physics, Chinese Academy of Sciences},
city={Lanzhou},
postcode={730000}, 
country={China}}

\begin{abstract}

The light nuclei yields and their yield ratios, regarded as sensitive probes of the QCD phase structure, have been extensively measured at various collision energies.
However, due to limited detector acceptance, the $p_{\rm T}$-integrated yield is often obtained by extrapolating from the measured $p_{\rm T}$ spectrum to the unmeasured low-$p_{\rm T}$ region using model-based fits.
Simulations using AMPT-HC combined with an after-burner coalescence approach indicate a significant enhancement of light nuclei production at low $p_{\rm T}$, particularly in peripheral collisions and at forward rapidities, driven primarily by spectator nucleons.
As a result, standard extrapolation procedures may systematically miss this additional low-$p_{\rm T}$ component, leading to an underestimate of the $p_{\rm T}$-integrated light-nucleus yields in such scenarios.
\end{abstract}


\end{frontmatter}




\section{Introduction}
\label{introduction}
One of the major goals of heavy-ion collision experiments is to study the Quantum ChromoDynamics (QCD) phase diagram~\cite{Ohnishi:2011aa, Palni:2024wdy}, which is often represented in terms of temperature vs. baryonic chemical potential. 
Lattice QCD calculations~\cite{Fodor:2004nz} and various theoretical models~\cite{Asakawa:1989bq, Stephanov:1998dy, Hatta:2002sj} indicate that the phase transition from Quark-Gluon Plasma (QGP) to hadronic matter is a smooth crossover at very small baryon chemical potential ($\mu_{\rm B}$), while at larger $\mu_{\rm B}$, it is likely to be a first-order phase transition. 
%
It is expected that low-energy collisions, which feature a large net-baryon density, will exhibit a first-order phase transition between the QGP and hadronic matter.
The production of light nuclei is an important observable in heavy-ion collision experiments, especially the double yield ratio (N$_p$ $\times$ N$_t$~/~N$^2$$_d$), which is considered a probe sensitive to the QCD phase transition~\cite{STAR:2022hbp}.
%
The production yields of light nuclei in heavy-ion collisions are commonly described by the thermal model and the coalescence model~\cite{Steinheimer:2012tb}.
%
According to the thermal model, the yield of light nuclei is significantly enhanced at RHIC fixed-target energies compared to the top energy. This enhancement can be well constrained by precise experimental measurements in the high baryon density region~\cite{ANDRONIC2011203,Donigus:2020fon}.
In addition, an approximate atomic mass number A scaling is shown in direct flow ($v_{1}$) of light nuclei, which is consistent with the nucleon coalescence model calculations~\cite{STAR:2021ozh,Han:2024xpg,Han:2024zyb}.
Recently, the RHIC-STAR collaboration published measurements of the yield and correlation of light nuclei at $\sqrt{s_{\rm{NN}}}$ = 3 GeV, which provide valuable insights into the production dynamics of light nuclei and the understanding of the QCD phase structure at high baryon density~\cite{STAR:2023uxk, STAR:2024zvj}.

Heavy-ion collisions are often described with a two-component picture.
In this framework, participants (or wounded nucleons) refer to the nucleons that undergo at least one inelastic interaction during the collision.
In contrast, spectators are the nucleons that do not participate in any inelastic scatterings and continue to travel essentially undeflected along the beam direction.
The two-component model assumes that the produced particles originate from a soft component that scales with the number of wounded nucleons ($\rm N_{\rm Part}$) and a hard component that scales with the number of binary nucleon-nucleon collisions ($\rm N_{\rm Coll}$)~\cite{Bialas:1976ed,Kharzeev:2000ph,Miller:2007ri,STAR:2008med}.
%
At high energies, such as those at the LHC and the RHIC top energy, the initial transverse momentum of nucleons (arising from Fermi motion within the nucleus) is on the order of $\sim$0.25 GeV. 
This is negligible compared to the longitudinal beam momentum per nucleon.
Therefore, whether the nucleons are wounded can be identified based on relative geometric position of nucleon-nucleon pairs from projectile and target nuclei.
This method is widely used as part of the initial collision conditions in model simulation of heavy-ion collision, such as heavy-ion jet interaction generator (HIJING)~\cite{GYULASSY1994307}, ultra-relativistic quantum molecular dynamics (UrQMD)~\cite{BASS1998255}, a multi-phase transport model (AMPT)~\cite{Lin:2004en}, etc.
%

However, when the collision energy decreases to a few GeV, corresponding to the high baryon density region, the motion of nucleons within the nuclei and their mutual interactions become significant and must be included in simulations.
In addition, as partonic interactions become weak or negligible, the evolution of the high-density nuclear matter created in heavy-ion collisions is likely dominated by hadronic interactions~\cite{STAR:2021yiu}.
Recently, a new version of the extended a multi-phase transport model (AMPT-HC) has been applied to study hadron cascade with baryonic mean-field potentials for heavy-ion collisions in the high baryon density region~\cite{Yong:2021npa,Wu:2023wfi,Wu:2023rui,Zhu:2025kud,Gao:2025voq}.
%
The literature cited above provides extensive discussions on the properties of yield and collective flow under the baryonic mean field.

In this paper, we investigate the spectator effect in light nuclei production in Au+Au collisions at $\sqrt{s_{\rm{NN}}}$ = 3 GeV, using the AMPT-HC model with an after-burner coalescence mechanism. Our results are compared with the corresponding experimental data from the STAR collaboration.
%
%
The paper is organized as follows. Firstly, the research backgrounds are introduced. Then, the details of model are described. Next, the simulation results are presented and discussed with experimental data. Finally, a simple summary is provided.

\begin{figure*}[h]
\centering
\includegraphics[width=0.95\textwidth,clip]{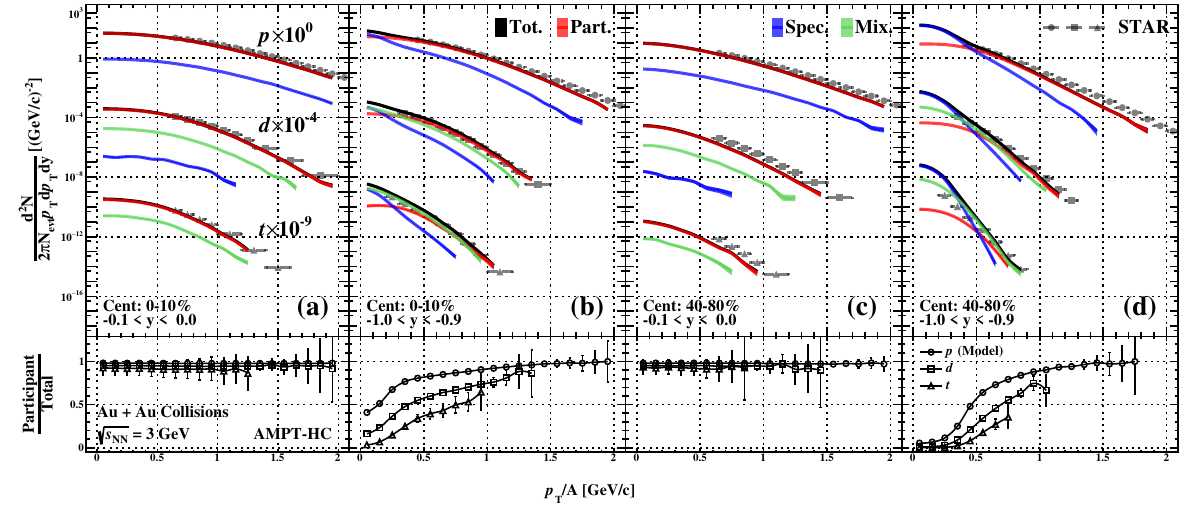}
\caption{Transverse momentum spectra of different components (top panels) and participant component fraction (bottom panels) of protons, deuterons, and tritons in Au+Au collisions at $\sqrt{s_{\rm{NN}}}$ = 3 GeV. The colored bands represent the AMPT-HC model simulations: black for the total component, red for the participant component, blue for the spectator component, and green for the mixing component. The solid marker points are STAR published data~\cite{STAR:2023uxk}. (a) centrality 0-10\% and mid-rapidity $-0.1 < y < 0$, (b) centrality 0-10\% and backward rapidity $-1.0 < y < -0.9$, (c) centrality 40-80\% and mid-rapidity $-0.1 < y < 0$, (d) centrality 40-80\% and backward rapidity $-1.0 < y < -0.9$.}
\label{Fig-1}
\end{figure*}

\section{Brief introduction of the model}
\label{model}

\subsection{Transport model}
\label{TransportModel}
The AMPT model is widely used to study heavy-ion collisions at high energies, such as LHC and RHIC top energies. 
It includes simulation of partonic and hadronic phase which consist of four main components: parton production from nucleons by HIJING, partonic interaction and transport by  Zhang’s parton cascade (ZPC)~\cite{ZHANG1998193}, hadronization by parton fragmentation or coalescence process, and hadronic interaction by a relativistic transport (ART)~\cite{Li:1995pra}.

The AMPT-HC model, a pure hadron cascade version, is designed to describe hadronic interaction in nucleus-nucleus collisions at center-of-mass energy of a few GeV. 
In this model, parton production, interaction and hadronization processes are switched off. 
The Woods-Saxon nucleon density distribution Eq.(\ref{Wood-Saxon}) is used to initialize the position of each nucleon within the nuclei.
\begin{gather}
    \label{Wood-Saxon}
    \rho(r)=\frac{\rho_{0}}{1+e^{-\frac{r-R}{d}}}
\end{gather}
Where $R$ is the half-density radius and $d$ is the diffuseness parameter, while $\rho_{0}$ denotes the normal nuclear density in Eq.(\ref{Wood-Saxon}).
And local Thomas-Fermi approximation Eq.(\ref{Thomas-Fermi}) is used to initialize the momentum.
\begin{gather}
    \label{Thomas-Fermi}
    p_{\rm F}(\rho)=0.197\times\left(\frac{3\pi^{2}\rho}{2}\right)^{1/3}
\end{gather}
Thomas-Fermi approximation depends on nuclear matter density, the coefficient 0.197 is the unit conversion between $\rm fm^{-1}$ and GeV. 
%
For example, at the center of a gold nucleus, the density $\rho \sim 0.168~\mathrm{fm}^{-3}$ corresponds to a Fermi momentum of $p_{\mathrm{F}} \sim 0.267~\mathrm{GeV}$.
Subsequently, the evolution of these nucleons is simulated by solving the Hamiltonian canonical equations (Eq.~\ref{HamiltonianEq}) in the presence of the fields they generate, including the electromagnetic field, kaon potentials~\cite{BROWN1994937}, baryon mean-fields~\cite{Gale19901545}, and others.
\begin{gather}
    \frac{d\vec{r}}{dt}=\frac{\partial H}{\partial \vec{p}} \notag \\
    \label{HamiltonianEq}
    \frac{d\vec{p}}{dt}=-\frac{\partial H}{\partial \vec{r}}
\end{gather}
%
The baryon mean-field, given by Eq.~(\ref{BaryonMeanField}), depends on the nuclear matter density $\rho$:
\begin{gather}
    U(\rho)=\alpha\frac{\rho}{\rho_{0}}+\beta\left(\frac{\rho}{\rho_{0}}\right)^{\gamma} \notag \\
    \alpha=\left(-29.81-46.9\frac{k_{0}+44.73}{k_{0}-166.32}\right)~\rm MeV \notag \\
    \beta=23.45\frac{k_{0}+255.78}{k_{0}-166.32}~\rm MeV \notag \\
    \label{BaryonMeanField}
    \gamma=\frac{k_{0}+44.73}{211.05}
\end{gather}
$k_{0}$ is called incompressibility which used to control stiffness of the baryon mean-field ($k_{0}$ is set to 380 MeV in this work~\cite{STAR:2021yiu,Gao:2025voq}). 
These particles, including mesons ($\pi$, $\rho$, $\omega$, $\eta$, $\rm{K}$, $\rm{K}^{*}$, $\phi$), baryons ($\rm{N}$, $\Delta$, $\rm{N}^{*}(1440)$, $\rm{N}^{*}(1535)$, $\Lambda$, $\Sigma$, $\Xi$, $\Omega$) and their anti-particles, can be produced or absorbed through inelastic collision channels during the transport process. 
In addition, the elastic collisions also occur between these particles, and the differential elastic scattering cross-section is described as below Eq.(\ref{EQ-1}),
\begin{gather}
    \label{EQ-1}
    \frac{d\sigma}{dt}=a\times e^{bt},~~t=-2p^{2}(1-\rm{cos}~\theta),
\end{gather}
where $p$ and $\theta$ represent momentum and scattering angle in center-of-momentum frame respectively. $a$ and $b$ depend on the collision energy and are taken from Ref.\cite{Bertsch:1988ik}.

Since Fermi motion and internucleon interactions are now included, the geometric criterion for determining the possibility of nucleon-nucleon inelastic scattering is no longer applicable.
By tracking the initial and final state of each nucleon-nucleon collision in the whole evolution of nucleus-nucleus collision, the nucleons that have undergone inelastic process will be recorded as participants, which is consistent with the Glauber model and the HIJING model~\cite{Miller:2007ri,GYULASSY1994307}, while the nucleons that have undergone elastic process or those that do not participate in collisions at all will be recorded as spectators.

\subsection{Coalescence model}
\label{CoalescenceModel}
Coalescence is commonly regarded as one of the primary mechanisms for the formation of light nuclei in heavy-ion collisions, and this process is generally performed after kinetic freeze-out in most models~\cite{Dover:1991zn}. 
When the transport evolution is completed, adjacent nucleons in the momentum and coordinate phase space can form light nuclei due to the final state interactions between nucleons.
The Wigner \cite{SUN2015272, Liu:2024ilw} and box \cite{Xu:2023xul, Sheikh:2022gbf} coalescence approaches are commonly used in heavy-ion collisions. In this work, we employ the box coalescence approach, as it can qualitatively describe the experimental data.
In this coalescence model, the relative position ${\rm \Delta} r$ and relative momentum ${\rm \Delta} p$ in the center-of-mass frame of a nucleon-nucleon pair are used to determine whether coalescence occurs~\cite{Steinheimer:2012tb, STAR:2022fnj,  Aichelin:2019tnk}.
%
Deuterons are formed via a ($p$, $n$) pair two-body coalescence model. For a $t$ formation, first a ($n$, $p$) or ($n$, $n$) pair is formed, and then an additional $n$ or $p$ is included.
%
Due to the strong short-range ($\lesssim 1$ fm) repulsion between nucleons~\cite{Day:1981zz}, we employ an elliptical window in phase space to determine nuclear formation.
\begin{gather}
    \label{EQ-coal}
    (\Delta r - r_{low})^2 / r_{high}^{2} + (\Delta p)^2 / p_{high}^{2} < 1
    ,~\Delta r > r_{low}
\end{gather}
Where the $r_{low}$ is the lower limit of $\Delta r$. The $r_{high}$ and $p_{high}$ are the long and short axes of the ellipse, respectively. They also denote the largest threshold that allows coalescence to occur.

The results of model calculations are obtained using the coalescence parameters for deuteron (triton) $r_{high}$ = 10 (8) fm , $p_{high}$ = 0.3 (0.2) GeV/$c$ and the $r_{low}$ = 0.8 (0.8) fm.
These parameters are obtained by minimizing the $\chi^{2}$ deviation Eq.(\ref{chi2test}) from the experimental data in the rapidity intervals $-0.1<y<0$ and $-1.0<y<-0.9$ for the 0–10\% and 40–80\% centrality classes. 
%
\begin{gather}
\label{chi2test}
\chi^{2} = \sum \frac{(N_{STAR} - N_{AMPT})^2}{\sigma^2_{STAR}}
\end{gather}
Here, $N_\text{STAR}$ and $N_\text{AMPT}$ denote the yields from the STAR experimental data and the AMPT calculation, respectively; $\sigma_\text{STAR}$ is the combined statistical and systematic uncertainty of the STAR data.

\section{Results and discussion}
\label{ResultsAndDiscussion}
In this work, we use the AMPT-HC model with an after-burner coalescence model to simulate protons, deuterons and tritons production in Au + Au collisions at $\sqrt{s_{\rm NN}}$ = 3 GeV.

\begin{figure}[h]
\centering
\includegraphics[width=8cm,clip]{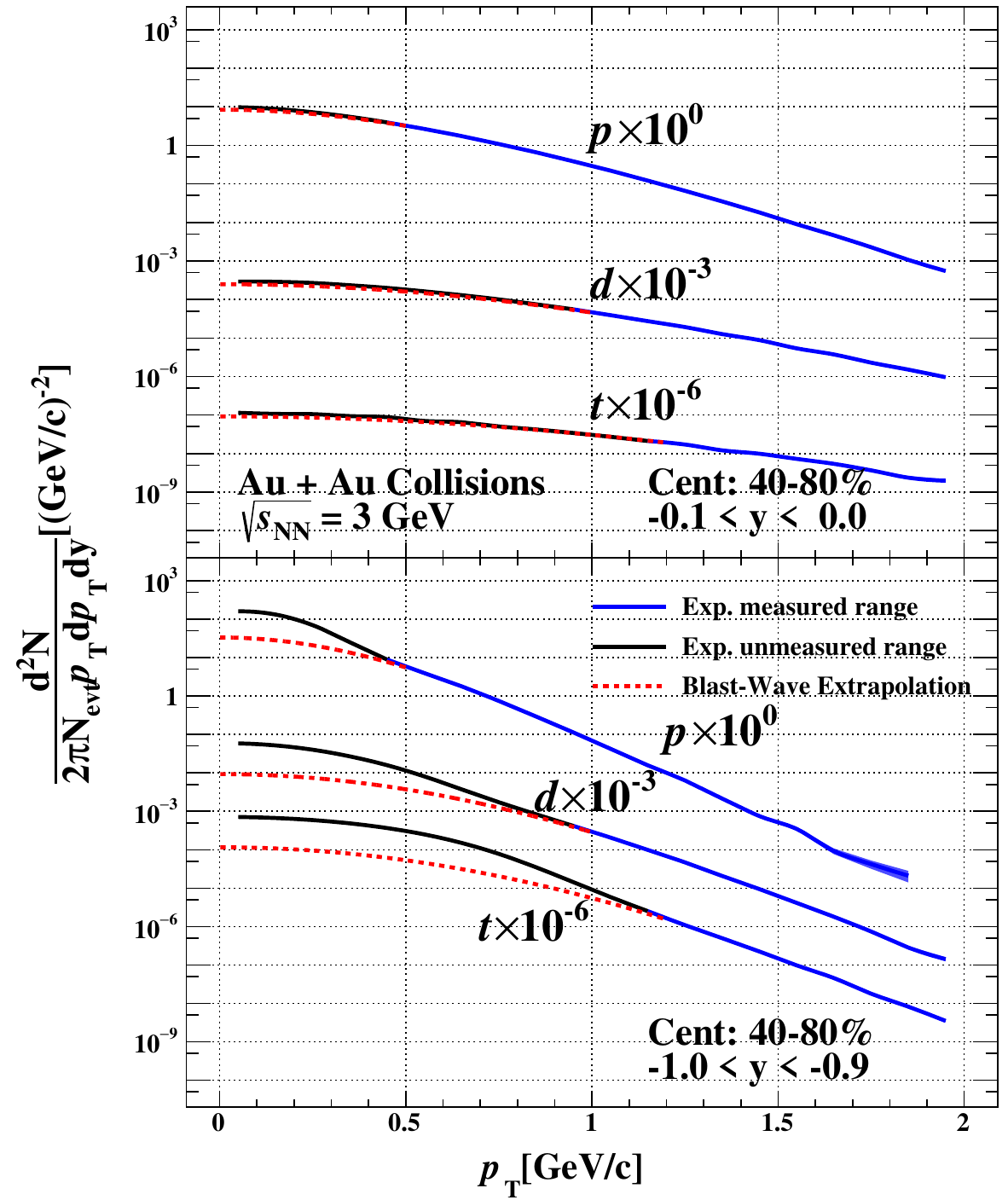}
\caption{Blast-Wave extrapolation of proton, deuteron, and triton spectra in peripheral Au+Au collisions (40–80\% centrality) at $\sqrt{s_{\rm NN}}=3$ GeV. Solid lines denote the AMPT-HC model simulations. The blue and black segments indicate the experimentally measured and unmeasured $p_{\rm T}$ ranges, respectively. The red dashed lines show the Blast-Wave extrapolation in the unmeasured $p_{\rm T}$ region, with the fit performed in the blue range.}
\label{Fig-2}
\end{figure}

In the four upper panels of figure~\ref{Fig-1}, the transverse momentum ($p_{\rm T}$) spectra of protons, deuterons, and tritons are presented for four rapidity and centrality bins: (a) mid-rapidity at 0-10\% centrality, (b) backward rapidity at 0-10\% centrality, (c) mid-rapidity at 40-80\% centrality, and (d) backward rapidity at 40-80\% centrality. The red and blue bands represent the participant component and the spectator component, respectively.
Deuteron and triton are formed through the coalescence of several nucleons. 
Therefore, they may contain nucleons from both participant and spectator sources, which is called mixed component as represented by green bands in the figure \ref{Fig-1}.
The black bands are the sum of all components, which represent the observable $p_{\rm T}$ spectra and it is called total $p_{\rm T}$ spectra, and the black solid markers represents the STAR measurements. 
The total $p_{\rm T}$ spectra of the AMPT-HC model (black bands) qualitatively reproduce the protons, deuterons and tritons of STAR measurements.
The $p_{\rm T}$ spectra of participant and spectator components in the AMPT-HC model show notably different shapes in different centrality and rapidity regions. 
%
%
At mid-rapidity (panel a and panel c), $p_{\rm T}$ spectra shapes between participant and spectator components are similar, while the spectator component is negligible.
Consequently, the fraction of participant component is near-unity, and shows no $p_{\rm T}$ dependence.
However, the $p_{\rm T}$ spectra of the spectator component show a much steeper slope compared with the participant component at backward rapidity.
Especially, in the low $p_{\rm T}$ range, such as $p_{\rm T} < 0.4$ GeV/$c$ for protons (panel b and panel d), the contribution from spectator component to the total yield is significantly enhanced.
This enhancement can be explained by the fact that spectator nucleons only suffer elastic scattering and carry the initial momentum, including both the longitudinal momentum from accelerated beam and the originated Fermi momentum within the nucleus. 
Therefore, in peripheral collisions or at backward rapidities, the spectator contribution to light-nuclei production becomes increasingly significant, leading to a noticeable modification of the $p_{\rm T}$ spectral shapes, particularly in the low-$p_{\rm T}$ region.

\begin{figure}
\centering
\includegraphics[width=9cm,clip]{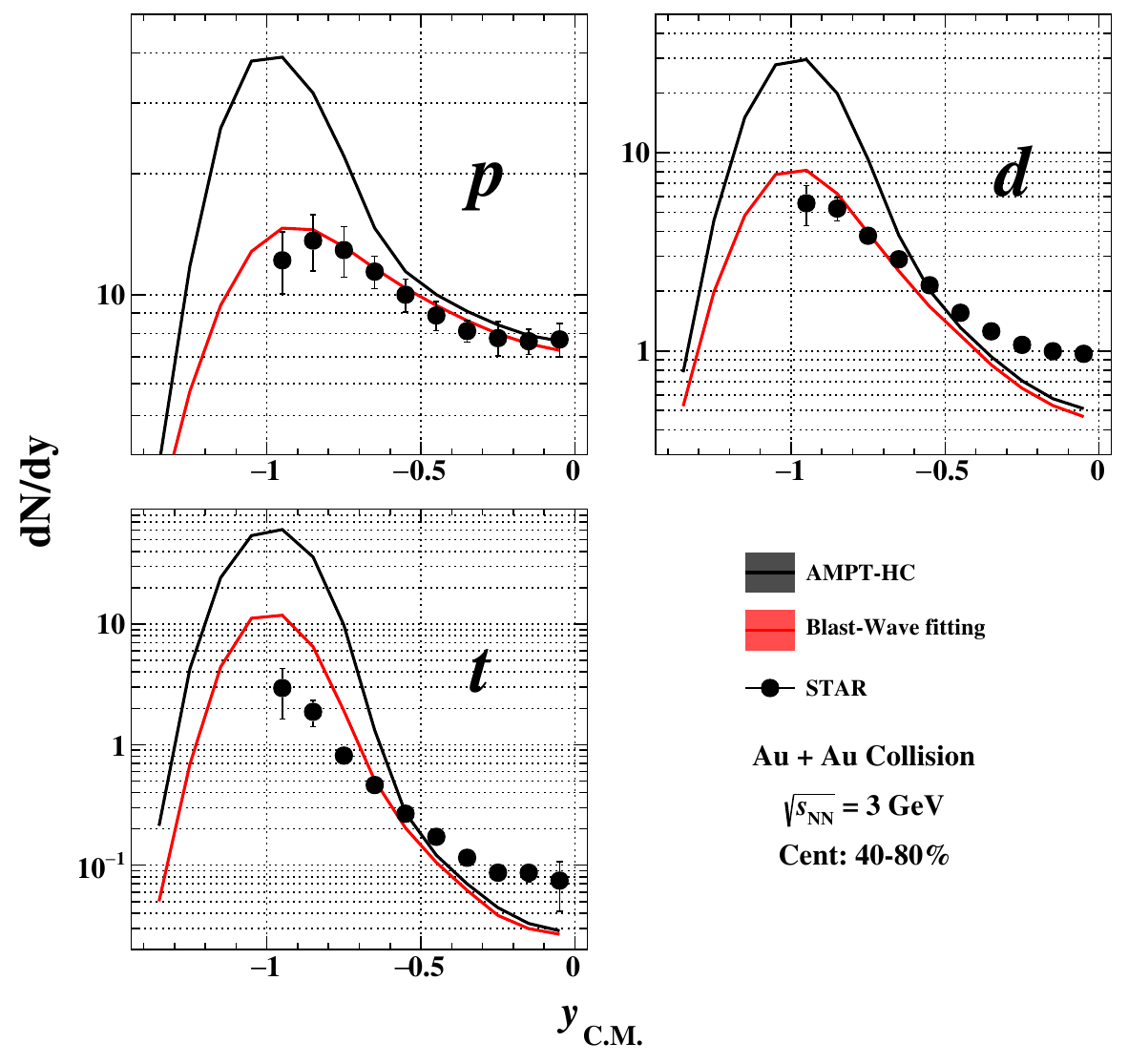}
\caption{Rapidity densities $dN/dy$ of protons, deuterons, and tritons in peripheral Au+Au collisions (40–80\% centrality) at $\sqrt{s_{\rm NN}}=3$ GeV. The black lines represent the integrals of the total $p_{\rm T}$ spectra over the full $p_{\rm T}$ range. The red lines are obtained by combining the integral of the Blast-Wave fit in the experimentally unmeasured $p_{\rm T}$ region with the integral of the AMPT-HC total $p_{\rm T}$ spectra in the experimentally measured $p_{\rm T}$ range. The black solid markers denote the published STAR data~\cite{STAR:2023uxk}.}
\label{Fig-3}
\end{figure}

In experiments, because of the limited detector acceptance, the $p_{\rm T}$ spectra are usually measured in a finite $p_{\rm T}$ range.
%
Therefore, in order to obtain the rapidity density distribution $d{\rm N}/dy$, a special function is used to fit the measured $p_{\rm T}$ spectrum, then the fitting function is used to extrapolate to the unmeasured $p_{\rm T}$ range, such as Eq.(\ref{ExperimentdNdy}), 
\begin{gather}
    \label{ExperimentdNdy}
    \frac{d{\rm N}}{dy} = \int_{unme.}
    f_{fit}(p_{\rm T})dp_{\rm T} + \int_{meas.}\left(\frac{d^{2}{\rm N}}{dp_{\rm T}dy}\right)_{Data}dp_{\rm T}
\end{gather}
and the Blast-Wave function is widely used in $p_{\rm T}$ spectra fitting and extrapolation~\cite{Schnedermann:1993ws, STAR:2024znc}.

We use the Blast-Wave function to fit middle and backward rapidity $p_{\rm T}$ spectra from the AMPT-HC model calculations under conditions, in peripheral collisions at 40-80\% centralities, assuming the limited $p_{\rm T}$ acceptance (experimental measured range) as fitting range: $p_{\rm T} > 0.5$ GeV/$c$ for proton, $p_{\rm T} > 1.0$ GeV/$c$ for deuteron and $p_{\rm T} > 1.2$ GeV/$c$ for triton.
The results are shown in Figure 2.
At middle rapidity, the Blast-Wave function successfully captures the $p_{\rm T}$ spectra beyond the fitted range. 
At backward rapidity, however, significant deviations between the AMPT-HC model and the Blast-Wave fit appear outside the fitted range, mainly due to a larger spectator component fraction in the low-$p_{\rm T}$ region than in the fitted region.

Figure~\ref{Fig-3} presents the rapidity density distribution $d{\rm N}/dy$ of proton, deuteron and triton in peripheral collisions at centrality 40-80\%.
The black line represents the integral of the total $p_{\rm T}$ spectra over the full $p_{\rm T}$ range, obtained by summing the integrals of the black and blue lines from figure~\ref{Fig-2}. 
The red line corresponds to the sum of two contributions: the integral of the Blast-Wave fitting function extrapolated into the experimentally unmeasurable $p_{\rm T}$ range (red line in figure~\ref{Fig-2}), and the integral of the total $p_{\rm T}$ spectra within the experimentally measurable $p_{\rm T}$ range (blue line in figure~\ref{Fig-2}).
The solid black markers represent the STAR measurements.
The $d{\rm N}/dy$ values predicted by the AMPT-HC model (black line) agree with the STAR data at mid-rapidity, but show a significant overestimation at backward rapidity. 
This discrepancy arises because the spectator component in the low \(p_{\mathrm{T}}\) range is not fully captured by extrapolation of the Blast-Wave function fit at backward rapidity.
As indicated by the red lines, the resulting $d{\rm N}/dy$ values become consistent with the STAR measurements across the entire rapidity range if the identical acceptance and fitting method are applied to the \(p_{\mathrm{T}}\) spectra generated by the AMPT-HC model.

\begin{figure}[h]
\centering
\includegraphics[width=8cm,clip]{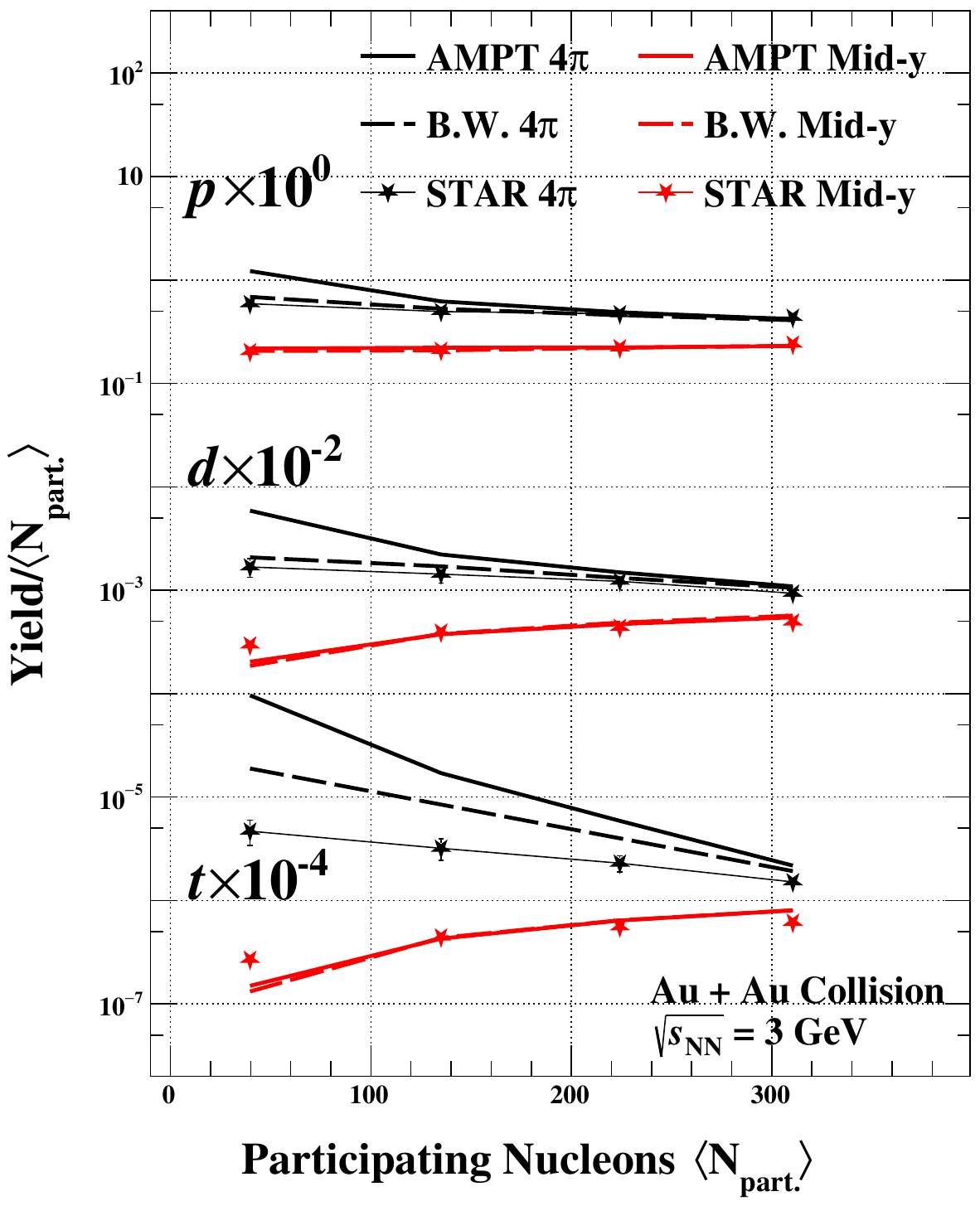}
\caption{Total (4$\pi$) and mid-rapidity ($-0.5 < y < 0$) yield as a function of the number of participating nucleons $\big<\rm N_{part.}\big>$ for each nucleon of $p, d$ and $t$ in Au+Au collisions at $\sqrt{s_{\rm{NN}}}$ = 3 GeV. The black and red colors represent the total and mid-rapidity yield respectively. The solid and dashed lines represent without or with Blast-Wave function extrapolation of $p_{\rm T}$ spectra. The star points are STAR published data~\cite{STAR:2023uxk}.}
\label{Fig-4}
\end{figure}

\begin{figure}[h]
\centering
\includegraphics[width=8cm,clip]{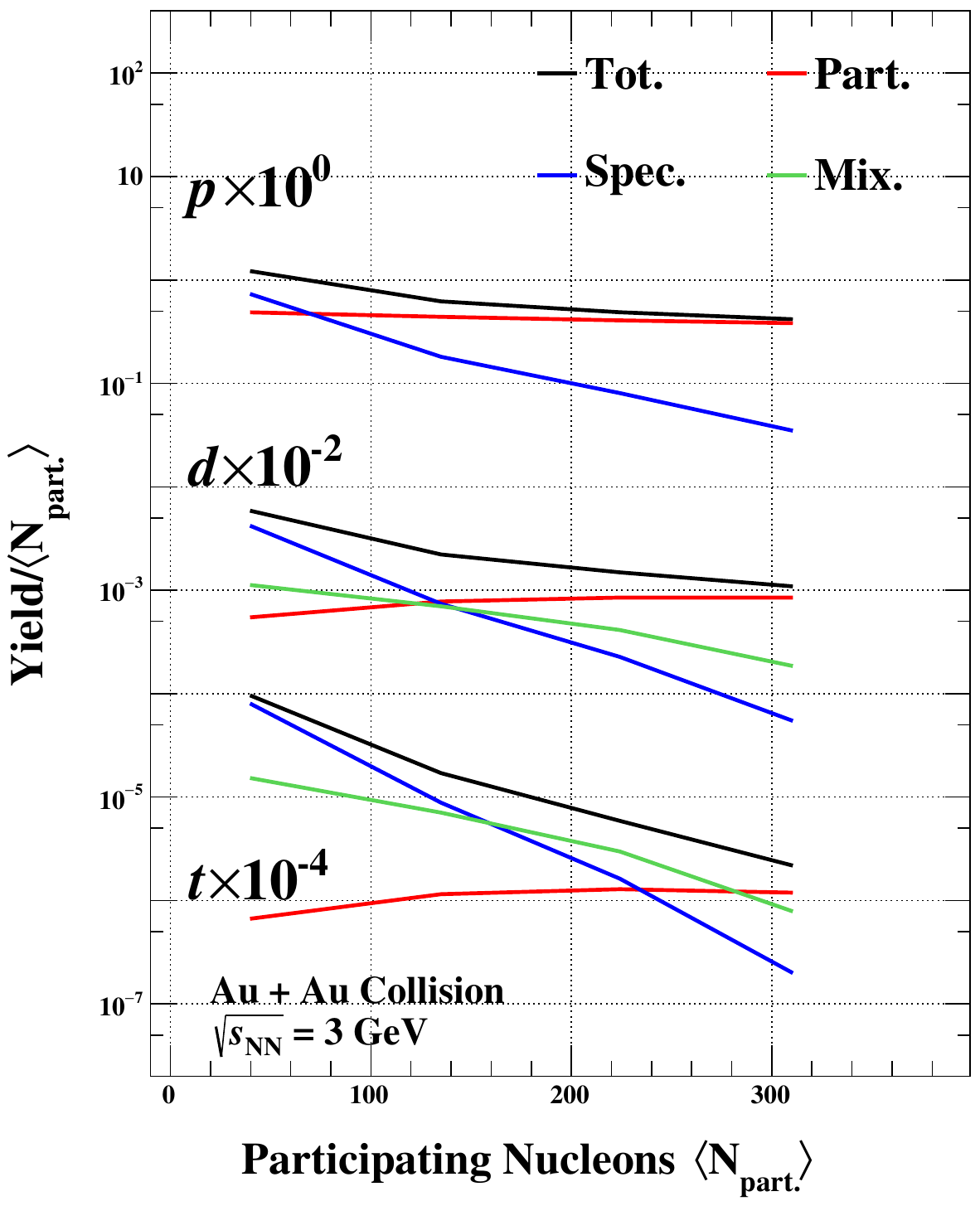}
\caption{Different components of total yield as a function of the number of participating nucleons $\big<\rm N_{part.}\big>$ for each nucleon of $p, d$ and $t$ in Au+Au collisions at $\sqrt{s_{\rm{NN}}}$ = 3 GeV. The black lines are total component, the red lines are participant component, the blue lines are spectator component, the green lines are mixing component.}
\label{Fig-5}
\end{figure}

Figure~\ref{Fig-4} shows the total and mid-rapidity yield as a function of the number of participant nucleons $\big<\rm N_{part.}\big>$ for protons, deuterons, and tritons in Au+Au collisions at \(\sqrt{s_{\mathrm{NN}}} = 3~\mathrm{GeV}\). 
The black and red colors represent total and mid-rapidity yield respectively. 
The solid and dashed lines represent without or with the extrapolation of Blast-Wave function fit for $p_{\rm T}$ spectra for the AMPT data. 
The star markers are from STAR published results\cite{STAR:2023uxk}.
The AMPT simulations describe well these trends that the mid-rapidity yields are increasing and total yields are decreasing from peripheral to central collisions.
The different centrality dependence between total and mid-rapidity yields arises from different components. 
The participant component contributes to the increasing behavior, while the spectator component contributes to the decreasing behavior, as shown in figure~\ref{Fig-5}. 
The figure~\ref{Fig-5} shows the different components of the total yield as a function of the number of participant nucleons $\big<\rm N_{part.}\big>$ for each nucleon of $p, d$ and $t$ in Au+Au collisions at $\sqrt{s_{\rm{NN}}}$ = 3 GeV, where the red, blue and green lines represent participant, spectator and mixing components, respectively.
In addition, as shown by the dashed lines in figure~\ref{Fig-4}, by using the same fitting extrapolation method as the STAR experiment, the centrality dependence of the total yield from the AMPT model will be closer to the experimental data. 
This further indicates that the extrapolation to the unmeasured low $p_{\rm T}$ range by the Blast-Wave function fitting is leading to a potential underestimation in conventional experimental yield calculations.
Our findings highlight the necessity of incorporating spectator effects in future measurements of light nuclei production, especially in the high-baryon-density region relevant to the QCD phase structure.
%

\section{Summary}
\label{Summary}
In summary, the AMPT-HC model with an after-burner coalescence reproduces the centrality dependence of light nuclei mid-rapidity and 4$\pi$ yield at $\sqrt{s_{\rm NN}}$ = 3 GeV.
Our calculations show that the spectator component contributes significantly to the low-$p_{\rm T}$ region, particularly in peripheral collisions and at forward rapidities, leading to a potential underestimation of yields in experimental extrapolations.
Incorporating the spectator effect is therefore essential for precise light nuclei yield determination in the high-baryon-density regime.

\section*{Acknowledgements}
This work is supported in part by the National Natural Science Foundation of China (Grant No. 12375134 and 12305146), the National Key Research and Development Program of China (Grant No. 2024YFE0110103 and 2024YFA1611003), the Fundamental Research Funds for the Central Universities (Grant No. CCNU25JCPT017), the Project for Young Scientists in Basic Research (No.YSBR-088) of the Chinese Academy of Science, and the Postdoctoral Fellowship Program of China Postdoctoral Science Foundation (Grant No. GZC20252245).


\bibliographystyle{elsarticle-num}
\bibliography{example}

@article{Ohnishi:2011aa,
    author = "Ohnishi, Akira",
    editor = "Asai, S. and Homma, Kensuke and Itakura, Kazunori and Nakamura, Atsushi and Shigaki, Kenta and Yamazaki, Yuji",
    title = "{Phase diagram and heavy-ion collisions: Overview}",
    eprint = "1112.3210",
    archivePrefix = "arXiv",
    primaryClass = "nucl-th",
    reportNumber = "YITP-11-106",
    doi = "10.1143/PTPS.193.1",
    journal = "Prog. Theor. Phys. Suppl.",
    volume = "193",
    pages = "1--10",
    year = "2012"
}

@article{STAR:2023uxk,
    author = "{STAR Collaboration}",
    collaboration = "STAR",
    title = "{Production of protons and light nuclei in Au+Au collisions at sNN=3 GeV with the STAR detector}",
    eprint = "2311.11020",
    archivePrefix = "arXiv",
    primaryClass = "nucl-ex",
    doi = "10.1103/PhysRevC.110.054911",
    journal = "Phys. Rev. C",
    volume = "110",
    number = "5",
    pages = "054911",
    year = "2024"
}

@article{Asakawa:1989bq,
    author = "Asakawa, M. and Yazaki, K.",
    title = "{Chiral Restoration at Finite Density and Temperature}",
    doi = "10.1016/0375-9474(89)90002-X",
    journal = "Nucl. Phys. A",
    volume = "504",
    pages = "668--684",
    year = "1989"
}

@article{Stephanov:1998dy,
    author = "Stephanov, Misha A. and Rajagopal, K. and Shuryak, Edward V.",
    title = "{Signatures of the tricritical point in QCD}",
    eprint = "hep-ph/9806219",
    archivePrefix = "arXiv",
    reportNumber = "ITP-SB-98-39, MIT-CTP-2748, SUNY-NTG-98-17",
    doi = "10.1103/PhysRevLett.81.4816",
    journal = "Phys. Rev. Lett.",
    volume = "81",
    pages = "4816--4819",
    year = "1998"
}

@article{Hatta:2002sj,
    author = "Hatta, Yoshitaka and Ikeda, Takashi",
    title = "{Universality, the QCD critical / tricritical point and the quark number susceptibility}",
    eprint = "hep-ph/0210284",
    archivePrefix = "arXiv",
    doi = "10.1103/PhysRevD.67.014028",
    journal = "Phys. Rev. D",
    volume = "67",
    pages = "014028",
    year = "2003"
}

@article{Fodor:2004nz,
    author = "Fodor, Z. and Katz, S. D.",
    title = "{Critical point of QCD at finite T and mu, lattice results for physical quark masses}",
    eprint = "hep-lat/0402006",
    archivePrefix = "arXiv",
    reportNumber = "ITP-BUDAPEST-609, WUB-04-04",
    doi = "10.1088/1126-6708/2004/04/050",
    journal = "JHEP",
    volume = "04",
    pages = "050",
    year = "2004"
}

@article{STAR:2022hbp,
    author = "{STAR Collaboration}",
    collaboration = "STAR",
    title = "{Beam Energy Dependence of Triton Production and Yield Ratio ($\mathrm{N}_t \times \mathrm{N}_p/\mathrm{N}_d^2$) in Au+Au Collisions at RHIC}",
    eprint = "2209.08058",
    archivePrefix = "arXiv",
    primaryClass = "nucl-ex",
    doi = "10.1103/PhysRevLett.130.202301",
    journal = "Phys. Rev. Lett.",
    volume = "130",
    pages = "202301",
    year = "2023"
}

@article{Palni:2024wdy,
    author = "Palni, Prabhakar and others",
    title = "{Dynamics of hot QCD matter 2024~{\textemdash} Bulk properties}",
    eprint = "2412.10779",
    archivePrefix = "arXiv",
    primaryClass = "nucl-th",
    doi = "10.1142/S0218301325440021",
    journal = "Int. J. Mod. Phys. E",
    volume = "34",
    number = "07",
    pages = "2544002",
    year = "2025"
}

@article{Donigus:2020fon,
    author = {D{\"o}nigus, Benjamin},
    title = "{Selected highlights of the production of light (anti-)(hyper-)nuclei in ultra-relativistic heavy-ion collisions}",
    doi = "10.1140/epja/s10050-020-00275-w",
    journal = "Eur. Phys. J. A",
    volume = "56",
    number = "11",
    pages = "280",
    year = "2020"
}

@article{Miller:2007ri,
    author = "Miller, Michael L. and Reygers, Klaus and Sanders, Stephen J. and Steinberg, Peter",
    title = "{Glauber modeling in high energy nuclear collisions}",
    eprint = "nucl-ex/0701025",
    archivePrefix = "arXiv",
    doi = "10.1146/annurev.nucl.57.090506.123020",
    journal = "Ann. Rev. Nucl. Part. Sci.",
    volume = "57",
    pages = "205--243",
    year = "2007"
}

@article{Yong:2021npa,
    author = "Yong, Gao-Chan and Xiao, Zhi-Gang and Gao, Yuan and Lin, Zi-Wei",
    title = "{Double strangeness {\ensuremath{\Xi}}{\ensuremath{-}} production as a probe of nuclear equation of state at high densities}",
    eprint = "2105.10284",
    archivePrefix = "arXiv",
    primaryClass = "nucl-th",
    doi = "10.1016/j.physletb.2021.136521",
    journal = "Phys. Lett. B",
    volume = "820",
    pages = "136521",
    year = "2021"
}

@article{Bertsch:1988ik,
    author = "Bertsch, G. F. and Das Gupta, S.",
    title = "{A Guide to microscopic models for intermediate-energy heavy ion collisions}",
    doi = "10.1016/0370-1573(88)90170-6",
    journal = "Phys. Rept.",
    volume = "160",
    pages = "189--233",
    year = "1988"
}

@article{GYULASSY1994307,
title = {HIJING 1.0: A Monte Carlo program for parton and particle production in high energy hadronic and nuclear collisions},
journal = {Computer Physics Communications},
volume = {83},
number = {2},
pages = {307-331},
year = {1994},
issn = {0010-4655},
doi = {https://doi.org/10.1016/0010-4655(94)90057-4},
url = {https://www.sciencedirect.com/science/article/pii/0010465594900574},
author = {Miklos Gyulassy and Xin-Nian Wang}
}

@article{BASS1998255,
title = {Microscopic models for ultrarelativistic heavy ion collisions},
journal = {Progress in Particle and Nuclear Physics},
volume = {41},
pages = {255-369},
year = {1998},
issn = {0146-6410},
doi = {https://doi.org/10.1016/S0146-6410(98)00058-1},
url = {https://www.sciencedirect.com/science/article/pii/S0146641098000581},
author = {S.A. Bass and M. Belkacem and M. Bleicher and M. Brandstetter and L. Bravina and C. Ernst and L. Gerland and M. Hofmann and S. Hofmann and J. Konopka and G. Mao and L. Neise and S. Soff and C. Spieles and H. Weber and L.A. Winckelmann and H. Stöcker and W. Greiner and Ch. Hartnack and J. Aichelin and N. Amelin}
}

@article{Lin:2004en,
    author = "Lin, Zi-Wei and Ko, Che Ming and Li, Bao-An and Zhang, Bin and Pal, Subrata",
    title = "{A Multi-phase transport model for relativistic heavy ion collisions}",
    eprint = "nucl-th/0411110",
    archivePrefix = "arXiv",
    doi = "10.1103/PhysRevC.72.064901",
    journal = "Phys. Rev. C",
    volume = "72",
    pages = "064901",
    year = "2005"
}

@article{ZHANG1998193,
title = {ZPC 1.0.1: a parton cascade for ultrarelativistic heavy ion collisions},
journal = {Computer Physics Communications},
volume = {109},
number = {2},
pages = {193-206},
year = {1998},
issn = {0010-4655},
doi = {https://doi.org/10.1016/S0010-4655(98)00010-1},
url = {https://www.sciencedirect.com/science/article/pii/S0010465598000101},
author = {Bin Zhang}
}

@article{Li:1995pra,
    author = "Li, Bao-An and Ko, Che Ming",
    title = "{Formation of superdense hadronic matter in high-energy heavy ion collisions}",
    eprint = "nucl-th/9505016",
    archivePrefix = "arXiv",
    doi = "10.1103/PhysRevC.52.2037",
    journal = "Phys. Rev. C",
    volume = "52",
    pages = "2037--2063",
    year = "1995"
}

@article{BROWN1994937,
title = {From kaon-nuclear interactions to kaon condensation},
journal = {Nuclear Physics A},
volume = {567},
number = {4},
pages = {937-956},
year = {1994},
issn = {0375-9474},
doi = {https://doi.org/10.1016/0375-9474(94)90335-2},
url = {https://www.sciencedirect.com/science/article/pii/0375947494903352},
author = {G.E. Brown and Chang-Hwan Lee and Mannque Rho and Vesteinn Thorsson}
}

@ARTICLE{Gale19901545,
	author = {Gale, C. and Welke, G.M. and Prakash, M. and Lee, S.J. and Das Gupta, S.},
	title = {Transverse momenta, nuclear equation of state, and momentum-dependent interactions in heavy-ion collisions},
	year = {1990},
	journal = {Physical Review C},
	volume = {41},
	number = {4},
	pages = {1545 – 1552},
	doi = {10.1103/PhysRevC.41.1545},
	url = {https://www.scopus.com/inward/record.uri?eid=2-s2.0-4243848848&doi=10.1103%2fPhysRevC.41.1545&partnerID=40&md5=377e18e6c6158e600425c4a7d8b1dfb8},
	type = {Article},
	publication_stage = {Final},
	source = {Scopus},
	note = {Cited by: 160}
}

@article{Schnedermann:1993ws,
    author = "Schnedermann, Ekkard and Sollfrank, Josef and Heinz, Ulrich W.",
    title = "{Thermal phenomenology of hadrons from 200-A/GeV S+S collisions}",
    eprint = "nucl-th/9307020",
    archivePrefix = "arXiv",
    reportNumber = "TPR-93-16",
    doi = "10.1103/PhysRevC.48.2462",
    journal = "Phys. Rev. C",
    volume = "48",
    pages = "2462--2475",
    year = "1993"
}

@article{STAR:2024znc,
    author = "{STAR Collaboration}",
    collaboration = "STAR",
    title = "{Strangeness production in $ \sqrt{s_{\textrm{NN}}} $ = 3 GeV Au+Au collisions at RHIC}",
    eprint = "2407.10110",
    archivePrefix = "arXiv",
    primaryClass = "nucl-ex",
    doi = "10.1007/JHEP10(2024)139",
    journal = "JHEP",
    volume = "10",
    pages = "139",
    year = "2024"
}

@article{STAR:2022fnj,
    author = "{STAR Collaboration}",
    collaboration = "STAR",
    title = "{Observation of Directed Flow of Hypernuclei H{\ensuremath{\Lambda}}3 and H{\ensuremath{\Lambda}}4 in sNN=3{\,}{\,}GeV Au+Au Collisions at RHIC}",
    eprint = "2211.16981",
    archivePrefix = "arXiv",
    primaryClass = "nucl-ex",
    doi = "10.1103/PhysRevLett.130.212301",
    journal = "Phys. Rev. Lett.",
    volume = "130",
    number = "21",
    pages = "212301",
    year = "2023"
}

@article{Steinheimer:2012tb,
    author = "Steinheimer, J. and Gudima, K. and Botvina, A. and Mishustin, I. and Bleicher, M. and Stocker, H.",
    title = "{Hypernuclei, dibaryon and antinuclei production in high energy heavy ion collisions: Thermal production versus Coalescence}",
    eprint = "1203.2547",
    archivePrefix = "arXiv",
    primaryClass = "nucl-th",
    doi = "10.1016/j.physletb.2012.06.069",
    journal = "Phys. Lett. B",
    volume = "714",
    pages = "85--91",
    year = "2012"
}

@article{Aichelin:2019tnk,
    author = "Aichelin, J. and Bratkovskaya, E. and Le F{\`e}vre, A. and Kireyeu, V. and Kolesnikov, V. and Leifels, Y. and Voronyuk, V. and Coci, G.",
    title = "{Parton-hadron-quantum-molecular dynamics: A novel microscopic $n$ -body transport approach for heavy-ion collisions, dynamical cluster formation, and hypernuclei production}",
    eprint = "1907.03860",
    archivePrefix = "arXiv",
    primaryClass = "nucl-th",
    doi = "10.1103/PhysRevC.101.044905",
    journal = "Phys. Rev. C",
    volume = "101",
    number = "4",
    pages = "044905",
    year = "2020"
}

@article{STAR:2024zvj,
    author = "{STAR Collaboration}",
    collaboration = "STAR",
    title = "{Light nuclei femtoscopy and baryon interactions in 3 GeV Au+Au collisions at RHIC}",
    eprint = "2410.03436",
    archivePrefix = "arXiv",
    primaryClass = "nucl-ex",
    doi = "10.1016/j.physletb.2025.139412",
    journal = "Phys. Lett. B",
    volume = "864",
    pages = "139412",
    year = "2025"
}

@article{ANDRONIC2011203,
title = {Production of light nuclei, hypernuclei and their antiparticles in relativistic nuclear collisions},
journal = {Physics Letters B},
volume = {697},
number = {3},
pages = {203-207},
year = {2011},
issn = {0370-2693},
doi = {https://doi.org/10.1016/j.physletb.2011.01.053},
url = {https://www.sciencedirect.com/science/article/pii/S0370269311001006},
author = {A. Andronic and P. Braun-Munzinger and J. Stachel and H. Stöcker}
}

@article{STAR:2021ozh,
    author = {STAR Collaboration},
    collaboration = "STAR",
    title = "{Light nuclei collectivity from~$\sqrt{s_{NN}}$ = 3 GeV Au+Au collisions at RHIC}",
    eprint = "2112.04066",
    archivePrefix = "arXiv",
    primaryClass = "nucl-ex",
    doi = "10.1016/j.physletb.2022.136941",
    journal = "Phys. Lett. B",
    volume = "827",
    pages = "136941",
    year = "2022"
}

@article{Han:2024xpg,
    author = "Han, Chengdong",
    collaboration = "STAR",
    title = "{Light- and hyper-nuclei collectivity in Au+Au collisions at RHIC-STAR}",
    doi = "10.1051/epjconf/202429605014",
    journal = "EPJ Web Conf.",
    volume = "296",
    pages = "05014",
    year = "2024"
}

@article{Han:2024zyb,
    author = "Han, Junyi",
    collaboration = "STAR",
    title = "{Directed Flow of {\ensuremath{\Lambda}}, {\ensuremath{\Lambda}}$^{3}$H and {\ensuremath{\Lambda}}$^{4}$H in Au+Au collisions at {\ensuremath{\sqrt{}}}sNN = 3.2, 3.5, 3.9 and 4.5 GeV at RHIC}",
    eprint = "2412.00871",
    archivePrefix = "arXiv",
    primaryClass = "nucl-ex",
    doi = "10.1051/epjconf/202531607008",
    journal = "EPJ Web Conf.",
    volume = "316",
    pages = "07008",
    year = "2025"
}

@article{Bialas:1976ed,
    author = "Bialas, A. and Bleszynski, M. and Czyz, W.",
    title = "{Multiplicity Distributions in Nucleus-Nucleus Collisions at High-Energies}",
    reportNumber = "TPJU-9/76",
    doi = "10.1016/0550-3213(76)90329-1",
    journal = "Nucl. Phys. B",
    volume = "111",
    pages = "461--476",
    year = "1976"
}

@article{Kharzeev:2000ph,
    author = "Kharzeev, Dmitri and Nardi, Marzia",
    title = "{Hadron production in nuclear collisions at RHIC and high density QCD}",
    eprint = "nucl-th/0012025",
    archivePrefix = "arXiv",
    doi = "10.1016/S0370-2693(01)00457-9",
    journal = "Phys. Lett. B",
    volume = "507",
    pages = "121--128",
    year = "2001"
}

@article{STAR:2008med,
    author = {STAR Collaboration},
    collaboration = "STAR",
    title = "{Systematic Measurements of Identified Particle Spectra in $p p, d^+$ Au and Au+Au Collisions from STAR}",
    eprint = "0808.2041",
    archivePrefix = "arXiv",
    primaryClass = "nucl-ex",
    doi = "10.1103/PhysRevC.79.034909",
    journal = "Phys. Rev. C",
    volume = "79",
    pages = "034909",
    year = "2009"
}

@article{Day:1981zz,
    author = "Day, B. D.",
    title = "{Three-body correlations in nuclear matter}",
    doi = "10.1103/PhysRevC.24.1203",
    journal = "Phys. Rev. C",
    volume = "24",
    pages = "1203--1271",
    year = "1981"
}

@article{Wu:2023wfi,
    author = "Wu, Zhi-Min and Yong, Gao-Chan",
    title = "{{\ensuremath{\Omega}}{\ensuremath{-}} production as a probe of the equation~of state of dense matter near the QCD phase transition in relativistic heavy-ion collisions}",
    eprint = "2307.06502",
    archivePrefix = "arXiv",
    primaryClass = "nucl-th",
    doi = "10.1103/PhysRevC.109.054903",
    journal = "Phys. Rev. C",
    volume = "109",
    number = "5",
    pages = "054903",
    year = "2024"
}

@article{Gao:2025voq,
    author = "Gao, Yuan and Zhang, Lei and Yu, Mei-Ling and Zheng, Hua and Yong, Gao-Chan",
    title = "{$\Xi^-$ collective flow and equation~of state of dense nuclear matter in heavy-ion collisions}",
    doi = "10.1103/b5zq-543n",
    journal = "Phys. Rev. C",
    volume = "112",
    number = "2",
    pages = "024909",
    year = "2025"
}

@article{Wu:2023rui,
    author = "Wu, Zhi-Min and Yong, Gao-Chan",
    title = "{Probing the incompressibility of dense hadronic matter near the QCD phase transition in relativistic heavy-ion collisions}",
    eprint = "2302.11065",
    archivePrefix = "arXiv",
    primaryClass = "nucl-th",
    doi = "10.1103/PhysRevC.107.034902",
    journal = "Phys. Rev. C",
    volume = "107",
    number = "3",
    pages = "034902",
    year = "2023"
}

@article{Zhu:2025kud,
    author = "Zhu, Xun and Yong, Gao-Chan",
    title = "{Violation of NCQ scaling in hadron elliptic flow in Au+Au collisions at sNN=3.0{\ensuremath{-}}7.7GeV}",
    eprint = "2505.07187",
    archivePrefix = "arXiv",
    primaryClass = "nucl-th",
    doi = "10.1016/j.physletb.2025.139752",
    journal = "Phys. Lett. B",
    volume = "868",
    pages = "139752",
    year = "2025"
}

@article{STAR:2021yiu,
    author = "{{STAR Collaboration}}",
    collaboration = "STAR",
    title = "{Disappearance of partonic collectivity in sNN=3GeV Au+Au collisions at RHIC}",
    eprint = "2108.00908",
    archivePrefix = "arXiv",
    primaryClass = "nucl-ex",
    doi = "10.1016/j.physletb.2022.137003",
    journal = "Phys. Lett. B",
    volume = "827",
    pages = "137003",
    year = "2022",
    note = "[Erratum: Phys.Lett.B 870, 139912 (2025)]"
}

@article{Dover:1991zn,
    author = "Dover, Carl B. and Heinz, Ulrich W. and Schnedermann, Ekkard and Zimanyi, Joszef",
    title = "{Relativistic coalescence model for high-energy nuclear collisions}",
    reportNumber = "BNL-45865, TPR-90-60",
    doi = "10.1103/PhysRevC.44.1636",
    journal = "Phys. Rev. C",
    volume = "44",
    pages = "1636--1654",
    year = "1991"
}

@article{SUN2015272,
title = {Production of antimatter 5,6Li nuclei in central Au+Au collisions at sNN=200 GeV},
journal = {Physics Letters B},
volume = {751},
pages = {272-277},
year = {2015},
issn = {0370-2693},
doi = {https://doi.org/10.1016/j.physletb.2015.10.056},
url = {https://www.sciencedirect.com/science/article/pii/S0370269315008151},
author = {Kai-Jia Sun and Lie-Wen Chen}
}

@article{Liu:2024ilw,
    author = "Liu, L. K. and Hu, C. L. and He, X. H. and Shi, S. S. and Xie, G. N.",
    title = "{Light and hyper nuclei formation at sNN= 3 GeV Au+Au collisions using Wigner coalescence approach}",
    eprint = "2404.13582",
    archivePrefix = "arXiv",
    primaryClass = "nucl-th",
    doi = "10.1016/j.physletb.2024.138853",
    journal = "Phys. Lett. B",
    volume = "855",
    pages = "138853",
    year = "2024"
}

@article{Xu:2023xul,
    author = "Xu, Yue and He, Xionghong and Xu, Nu",
    title = "{Light nuclei production in Au+Au collisions at 3 GeV from coalescence model*}",
    eprint = "2305.02487",
    archivePrefix = "arXiv",
    primaryClass = "nucl-th",
    doi = "10.1088/1674-1137/acd3d9",
    journal = "Chin. Phys. C",
    volume = "47",
    number = "7",
    pages = "074107",
    year = "2023"
}

@article{Sheikh:2022gbf,
    author = "Sheikh, Ashik Ikbal",
    title = "{Investigating the coalescence-inspired sum rule for light nuclei and hypernuclei in heavy-ion collisions}",
    eprint = "2206.14296",
    archivePrefix = "arXiv",
    primaryClass = "nucl-th",
    doi = "10.1103/PhysRevC.106.054907",
    journal = "Phys. Rev. C",
    volume = "106",
    number = "5",
    pages = "054907",
    year = "2022"
}


\end{document}